\documentclass[twocolumn,showpacs,preprintnumbers,amsmath,amssymb]{revtex4}

\usepackage{graphicx}
\usepackage{dcolumn}
\usepackage{bm}

\begin{document}

\preprint{APS/123-QED}

\title{Deformed Special Relativity with a minimum speed as explanation of the tiny value of the cosmological constant based on the Boomerang experiment in the $\Lambda CDM$ scenario}  
\author{Cl\'audio Nassif Cruz}
\altaffiliation{{\bf UFOP}: Universidade Federal de Ouro Preto, Morro do Cruzeiro, Bauxita, 35.400-000-Ouro Preto-MG, Brazil \\
email: claudionassif@yahoo.com.br}
\author{A. C. Amaro de Faria Jr}
\altaffiliation{{\bf UTFPR-GP}: Federal Technological University of Paran\'a, Av. L. Bastos, 85053-525 , Guarapuava-PR, Brazil (permanent) \\
{\bf $^\dagger$IEAv}: Advanced Studies Institute-IEAv, Trevo Cel. Amarante,  12.228-001, Sao Jose dos Campos-SP, Brazil\\
email: atoni.carlos@gmail.com} 



\date{\today}

\begin{abstract}

In this paper we will show that a new structure of space-time with an invariant minimum speed reveals a connection with Weyl geometry in the approximation of weak-field Newtonian limit. Symmetrical Special Relativity (SSR) has an invariant minimum speed $V$ that plays the role of a preferred reference frame $S_V$ of vacuum that leads to the cosmological constant $\Lambda$. In order to realize such a connection between $V$ and $\Lambda$ within a scenario of space-time metric, we will use a model of spherical universe with Hubble radius $R_H$ filled by a low vacuum energy density $\rho_{\Lambda}$ that governs the accelerated expansion of the universe. In doing this, we will show that SSR-metric plays the role of a de-Sitter (dS)-metric with a positive cosmological constant ($\Lambda>0$). On the other hand, according to the Boomerang experiment as it is shown that the three-dimensional space of the universe is Euclidean and with a slightly accelerated expansion, SSR leads to a dS-metric with an approximation for $\Lambda<<1$ close to a flat space-time, which is in the $\Lambda CDM$ scenario where the space is quasi-flat, so that $\Omega_{m}+\Omega_{\Lambda}\approx 1$. We have $\Omega{cdm}\approx 23\%$ by representing dark cold matter, $\Omega_m\approx 27\%$ for matter and $\Omega_{\Lambda}\approx 73\%$ for the vacuum energy. Thus, the theory is adjusted for redshift $z=1$, i.e., the time $\tau_0$ at which the universe goes over from a decelerating to an accelerating expansion by obtaining the numerical value $\Lambda_0=1.934\times 10^{-35}s^{-2}$, being in good agreement with measurements. 

\end{abstract}

\maketitle

\section{\label{sec:level1} Introduction}

In 1905, Einstein published his paper on Special Relativity (SR), entitled 
{\it ``On the Electrodynamics of Moving Bodies''}\cite{Einstein}, where he changed the laws of Newtonian mechanics in order to preserve the covariance of Maxwell equations, so that the speed of light in vacuum ($c\cong 2.99792458\times 10^{8}$m/s) must mantain invariant for any inertial motion. Thus $c$ is the maximum limit of speed, which is unattainable for any massive particles, except the photon with speed $c$, as it is a massless particle. In view of this, the space, time, mass and energy become related between themselves, as all these quantities depend on speed. However, SR was built on an empty space, i.e., there is no kind of {\it aether} or no vacuum energy in SR, as the uncertainty principle (the zero-point energy) is out of the structure of space-time of SR. The great challenge is the natural inclusion of the quantum principles associated with a fundamental vacuum energy into a new structure of space-time, where the cosmological constant emerges naturally from such first principles that should be investigated.  

We will search for a new structure of space-time with the presence of a minimum speed $V$ that behaves like a kinematic invariant for particles with low energies as is the speed of light $c$ for high energies, by forming a fundamental symmetry of motion that should be justified by first principles, which must be consistent with the quantum principle concerning the zero-point energy as an effect of the own uncertainty principle.

Such zero-point energy is associated with the vacuum energy that leads to the cosmological constant. {\it In this sense, wouldn't it be natural to realize that the classical conception of rest idealized in the quantum world is not in fact compatible with the zero-point energy, even more because the zero-point energy has gravitational origin so that the particle is not totally free of gravity by forming a bound state with the whole universe, as gravity is everywhere}? If it is so, then we intend to build a modified relativity in order to become compatible with such vacuum energy or zero-point energy, so that we would be motivated to postulate a new kinematic invariance for low energies, i.e., an invariant minimum speed $V$ in such a new structure of space-time to be better explored and justified later. Thus we have a kind of quantum space-time with two kinematic invariants $c$ and $V$. 

Actually, it is important to stress that $V$ must be invariant because there would be no referential that nullify it, otherwise we would be returning to the classical concept of rest, which is not allowed in this quantum space-time. Such invariance of $V$ will be shown later by means of new velocity transformations, where the invariance of $V$ is represented
by a preferred (universal) reference frame $S_V$ given by a cosmic background field (vacuum energy) as explanation for the cosmological constant. 

In sum, we will be led to a better understanding of the cosmological implication with respect to the own cosmological constant within a kinematic scenario described by a modified relativity with an 
invariant minimum speed associated with a preferred reference frame given by vacuum energy. 

So, finally we are led to think that such symmetry due to $c$ and $V$ forms the kinematic basis in a space-time that bahaves like a de-Sitter (dS) space-time represented by a positive cosmological constant $\Lambda$. In view of such a connection between Deformed Special Relativity (DSR) with a minimum speed and a dS space-time, we can get a tiny value of $\Lambda$, which is associated with a weak cosmological anti-gravity, such that we will find $\Lambda=\Lambda_0=1.934\times 10^{-35}s^{-2}$ for $z=1$ in the zero gravity-limit when the anti-gravity begins to emerge, i.e., the accelerated cosmic expansion comes into play.

The search for understanding the origin of the vacuum energy density $\rho_{\Lambda}$ related to $\Lambda>0$ in the scenario of an accelerated expanding universe has been the issue of hard investigations\cite{1}\cite{2}, where it is known that the vacuum energy density is $\rho_{\Lambda}=\Lambda c^2/8\pi G$. 

The fine structure constant $\alpha$ is associated with the cosmological constant $\Lambda$\cite{hgn,hgn2,hgn3}. Thus, a possible variation of the fine structure constant\cite{sergio,sergio1,sergio2,pad,pad2} would also point to a fundamental change in the subatomic structure, since $\alpha$ has a property of connecting the micro and macro-world, whose age is measured by the speed of light $c$, i.e., $R_H=cT_H$, where $T_H(\cong 13.7$ Gyear) is the Hubble time and $R_H(\sim 10^{26}m)$ is the Hubble radius, i.e., the radius of the visible universe.  

The relationship between $\alpha$ and $\Lambda$, which is linked to the dark sector of the universe is associated with the models that aim to explain the anti-gravitational effects of the dark energy based on scalar fields\cite{cine1,cine2,cine3,cine4,cine5,cine6,
cine7}. 

The emergence of a minimum speed $V$ in the
space-time is associated with a preferred reference frame, thus leading to the birth of a new relativity with Lorentz symmetry violation at lower energies, i.e., the so-called Symmetrical Special Relativity (SSR)\cite{N2016,N2012,N2010,Rodrigo,Rodrigo2,Rodrigo3,N2018,uncertainty}.

It has also been shown that SSR has a relationship with the principle of Mach\cite{Rodrigo3,mach,mach2,mach3} within a quantum scenario due to the presence of the vacuum energy\cite{Rodrigo}.

There is a relationship between the fine structure constant and the cosmological constant, i.e., $\Lambda\propto\alpha^{-6}$\cite{hgn,hgn2,hgn3}. Actually, 
$\Lambda$ is also associated with other constants such as the mass of the electron $m_{e}$, Planck constant ($\hbar$) and the universal constant of gravity $G$. Thus, we can realize that $\Lambda$ is connected to the constants of the standard model of elementary particles, namely 
$\Lambda\sim (G^2/\hbar^4)(m_e/\alpha)^6$\cite{hgn,hgn2,hgn3}.

The third section will be dedicated to the introduction of the space-time and velocity transformations for ($1+1$)D in SSR-
theory\cite{N2016,N2012,N2010,Rodrigo,Rodrigo2,Rodrigo3}.

In the section 4, our goal is to show that the SSR-metric plays the role of a dS-metric, so that $\Lambda$ emerges naturally from the SSR-theory 
by using a simple model of spherical universe with Hubble radius filled 
by a uniform vacuum energy density $\rho_{\Lambda}$. Thus, $\Lambda$ is related to a cosmological anti-gravity. The tiny order of magnitude of $\Lambda(\sim 10^{-35}s^{-2})$ can be estimated. 

In the subsection 4.1, by making the approximation for a very weak 
anti-gravity in the dS-metric from SSR, i.e., $\Lambda<<1$, we are
within a more realistic cosmological scenario of a slightly accelerated expanding quasi-flat space-time, according to the observational data provided by the Boomerang experiment. So, we can go even further by getting the tiny numerical value of 
$\Lambda=1.934\times 10^{-35}s^{-2}$ given by the observations at the redshift $z=1$ so-called zero-gravity limit when the universe goes over from a decelerating to an accelerating expansion\cite{boomerang}\cite{boomerang1}. 

Finally, the section 5 is dedicated to the Weyl 
geometrical structure of SSR. In the Weyl scenario of conformally flat spacetimes, we will show in a simple and direct way that the
factor $\Theta(v)$ in Eq.(3) of SSR behaves like a conformal Weyl factor so that SSR includes a Weyl conformal geometry in the regime of Newtonian 
weak-field, i.e., for $\phi<<1$ by considering 
$c=1$, such that $\Theta\cong 1$, which is the own conformal factor of SSR given for 
the weak-field regime, where the space-time is almost flat. 

The great relevance of this regime of weak-field in the Weyl structure is that such regime corresponds to the slight acceleration of the universe for 
$z=1$, where we are able to get the tiny value of the cosmological constant according to the experiments. So we will conclude that the actual 
expanding universe is governed by a Weyl conformal geometry for weak-field by representing an almost flat space-time as a particular case of Eq.(3). 

It is important to notice that the Weyl structure was originally proposed with the aim of presenting a unification model between Electromagnetism and Gravitation. The purpose and importance of this work is to show that the Weyl structure emerges from the SSR at the weak field boundary and from which we can obtain the tiny value of the cosmological constant. Specifically speaking, the Weyl field is responsible for the conformal structure of the theory related to a quasi-flat space-time metric. In this sense, we show that the SSR conformal factor $\Theta$, in addition to being conformally flat, is directly related to the Weyl factor with the same approximation in the weak field limit,
i.e, $\Theta\approx 1$ ($\phi<<1$). This result is important as it shows that the Minkowski space-time metric can be slightly perturbed showing that the Weyl structure, to some extent, manifests itself in the weak-field boundary of the SSR. In the cosmological scenario, such weak-field regime ($\Theta\approx 1$) occurs 
for the galaxies with redshift $z=1$ when occurred the transition from gravity to anti-gravity with a slight acceleration represented by a very small positive cosmological constant to be obtained according to the observational data of the Boomerang experiment.

\section{\label{sec:level1} The hypothesis of a minimum speed in the space-time} 

The motivation for considering the existence of a lowest non-null limit of speed for very low energies ($v<<c$) in the space-time results in the following physical reasoning: 

- In non-relativistic Quantum Mechanics (QM), the plane wave wave-function ($Ae^{\pm ipx/\hbar}$), which represents a free particle is an idealization that is impossible to conceive under physical reality unless we make some approximations just for practical purposes. In the event of such an idealized plane wave, it would be possible to find with certainty the reference frame that cancels its momentum ($p=0$), so that the uncertainty on its position would be infinite or simply diverge ($\Delta x=\infty$). However, the hypothesis of the existence of a minimum limit of speed $V$ in the space-time avoids such unrealistic extreme condition of a perfect plane wave in QM, since $V$ emerges in order to prevent this ideal case of a plane wave ($\Delta p=0$), where the uncertainty on position diverges. In other words, we can realize that the existence of a minimum speed works like a cut-off for lower speeds by avoiding the existence of rest, which leads to a realistic condition where there is no perfect plane wave in reality, except the own preferred reference frame associated with the minimum speed $V$ to be postulated as an invariant speed, since the momentum of the particle tends to zero when the speed $v$ is so close to $V$, but it never reaches $V$. 

Furthermore, here it is important to point out that the infinite uncertainty on position of the particle in its plane wave with a well-defined momentum is in fact an idealization, since it is not in agreement with the cosmological reality of a finite universe, whose Hubble radius $R_H\sim 10^{26}$m leads us to think of a maximum uncertainty on position of a particle, but having a finite order of magnitude due to the own finite radius of the universe, as if the particle were free inside a big box
($\Delta x\sim 10^{26}$m), thus having a quasi-zero minimum  uncertainty on momentum ($\Delta p_{min}>0$). This justifies the existence of a minimum speed in the space-time of a deformed relativity to be investigated. Therefore we can have a quasi-plane wave, but never a perfect plane wave with 
$\Delta x=\infty$, since the radius of the universe is finite.  
Thus, as the universe is finite, we can think about a simple model of a particle inside a box (universe) with the order of magnitude of $10^{26}$m instead of the ideal case of a plane wave for a free particle ($\Delta x=\infty$) with a null zero-point energy, which is prevented by the minimum speed $V$, being consistent with the realistic case of a particle inside
a finite box by representing a finite universe with a non-
null zero-point energy. 

Due to the finite radius of the universe, such a particle has a non-null zero-point energy, which is in agreement with the 
impossibility of rest emerging from the uncertainty principle. In this sense, we can realize that the absence of rest justified by the minimum speed $V$ further clarifies the understanding of the own uncertainty principle\cite{N2012}, which already establishes a non-null zero-point energy associated with the vacuum energy in the cosmological scenario. 

As the vacuum energy has a quantum-gravitational origin due to a fundamental zero-point energy related to a universal minimum speed $V$, we expect there should be a relationship between 
$V$ and gravity ($G$), as it was shown in a previous paper, i.e., it was found $V\sim G^{1/2}$\cite{N2016}, $G$ being the constant of gravity.   

The existence of a minimum speed $V$ reveals us that the luminal particles as the photon ($v=c$) as well as the massive or subluminal particles ($v<c$) are in equal-footing in the sense that it is not possible to find a reference frame at rest ($v_{relative}=0$) for any velocity transformations in the space-time with both maximum and minimum speed limits so-called Symmetrical Special Relativity (SSR). Such transformations will be shown in the next section. 

The dynamics of particles in the presence of a universal background reference frame associated with the minimum speed
$V$ strengthens the basic idea provided by the scenarios 
of Mach\cite{Mach4}, Schr\"{o}dinger\cite{Schroedinger} and Sciama\cite{Sciama}, where there should be an absolute inertial reference frame in relation to which we have the inertia of all moving bodies. However, we must emphasize that the approach used here is not classical as the machian ideas, since the unattainable minimum speed $V$ has quantum origin by playing the role of a preferred reference frame of background field (vacuum energy) instead of the ``inertial'' frame of the apparent fixed stars. 

\section{\label{sec:level1} Space-time and velocity transformations in SSR} 

In this section, first of all we will carefully investigate the concepts of reference frame in SSR and their implications such as the new transformations of space-time and velocity.

In the next section, we will show the equivalence between the SSR-metric and a dS-metric, which is associated with a certain positive cosmological constant $\Lambda(>0)$, thus leading to a cosmological anti-gravity. So, we will conclude that SSR generates a background metric that plays the role of a dS-metric with the presence of $\Lambda$. In order to show such a fundamental equivalence, we will use a toy model, as it will be presented in the next section. 

Violation of Lorentz symmetry for very low energies\cite{N2016} generated by the presence of a background field related to $S_V$ (Fig.1) creates a space-time with an invariant mimimum speed 
$V(=\sqrt{Gm_pm_e}e/\hbar\sim 10^{-14}m/s$)\cite{N2016}, which is the unattainable limit of speed for all particles at lower energies in SSR. 

Since the minimum speed $V$ is an invariant quantity as is the speed of light $c$, $V$ does not alter the speed $v$ of any particle, as we will show later. Therefore, we denominate ultra-referential $S_V$ as being the preferred reference frame in relation to which we have the speeds $v$ of any particles (Fig.1). In view of this, the well-known Lorentz transformations are changed in the presence of the background reference frame $S_V$ (Fig.1). 

In the case $(1+1)D$, we have obtained the following space-time transformations\cite{N2016}\cite{N2012} given between the running reference frame $S^{\prime}$ and the background reference frame $S_V$ (Fig.1), namely: 

\begin{figure}
\begin{center}
\includegraphics[scale=0.087]{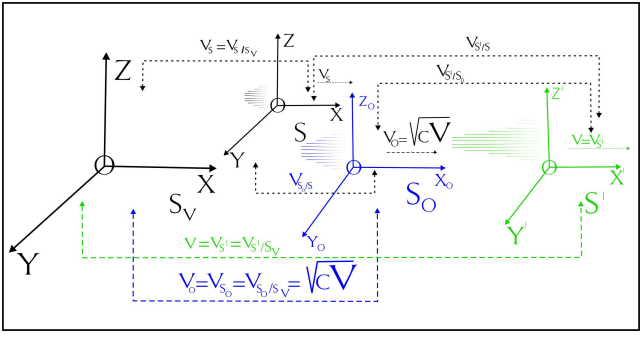} 
\end{center}
\caption{The reference frame $S^{\prime}$ moves in $x$-direction with a speed $v(>V)$ with respect to the universal reference frame, i.e., the ultra-referential $S_V$ associated with $V$. In this figure, we see the two running referentials $S$ and $S^{\prime}$ with speeds $v=v_S=v_{S/S_V}$ and $v^{\prime}=v_S'=v_{S'/S_V}$, both of them given in relation to the background frame (ultra-referential $S_V$), plus two fixed referentials $S_0$ with speed $v_0=v_{S_0/S_V}=\sqrt{cV}$ given with respect to the background frame $S_V$ and the own ultra-referential $S_V$ of vacuum associated with the unattainable minimum speed $V$. Thus we can find the relative velocity between $S'$ and $S$, i.e., $v_{rel}=v_{S'/S}$, which is shown clearly in Eq.(5) and Eq.(6), thus leading to some important cases, as for instance: a) If only the running referential $S$ coincides with $S_0$ ($S\equiv S_0$), we find the relative velocity between $S'$ and $S_0$, i.e., $v_{rel}=v_{S'/S_0}$. b) If only the running referential $S'$ coincides with $S_0$ ($S'\equiv S_0$), we find the relative velocity between $S_0$ and $S$, i.e., $v_{rel}=v_{S_0/S}$, as it is also indicated in this figure.} 
\end{figure}

\begin{equation}
dx^{\prime}=\frac{\sqrt{1-V^2/v^2}}{\sqrt{1-v^2/c^2}}[dX-v(1-\alpha)dt]
\end{equation}

and 

\begin{equation}
 dt^{\prime}=\frac{\sqrt{1-V^2/v^2}}{\sqrt{1-v^2/c^2}}\left[dt-\frac{v(1-\alpha)dX}{c^2}\right], 
 \end{equation}
with $\alpha=V/v$ and $\Psi=\theta\gamma=\sqrt{1-V^2/v^2}/\sqrt{1-v^2/c^2}$, where $\theta=\sqrt{1-V^2/v^2}$ and $\gamma=1/\sqrt{1-v^2/c^2}$. 

The coordinates $X$ shown in the transformations $(1+1)D$ above, $Y$ and $Z$ (Fig.1) form the ultra-referential $S_V$ connected to the vacuum energy. 

The inverse transformations for this special case $(1+1)D$ (Fig.1) were demonstrated in a previous work\cite{N2016}. Of course, if we make $V\rightarrow 0$, we recover the Lorentz transformations. 

The general transformations in the space-time $(3+1)D$ of SSR were also shown in a previous paper\cite{N2016}. In this previous paper\cite{N2016}, it was shown that SSR transformations breaks down the Lorentz and Poincar\'e's groups.

This new causal structure of space-time, i.e., the Symmetrical Special Relativity (SSR) presents the following energy $E$ and momentum $P$ for a particle, namely $E=m_0c^2\Psi=m_0c^2\sqrt{1-V^2/v^2}/\sqrt{1-v^2/c^2}$\cite{N2016}, in such a way that $E\rightarrow 0$ when $v\rightarrow V$, and $P=m_0v\Psi=m_0v\sqrt{1-V^2/v^2}/\sqrt{1-v^2/c^2}$\cite{N2016}, such that $P\rightarrow 0$ when $v\rightarrow V$. 

It is important to notice that both momentum-energy of a particle in SSR is $P_0=m_0v_0=m_0\sqrt{cV}$ and $E_0=E(v_0)=m_0c^2$ for $v=v_0=\sqrt{cV}\neq 0(>V)$, as we find $\Psi(v_0)=\Psi(\sqrt{cV})=1$, where the energy $m_0c^2$ is exactly equivalent to the rest energy in SR, since there is no rest in SSR. This means that the momentum never vanishes in SSR due to the invariant minimum speed $V$, as there is an intermediary speed $v_0(=\sqrt{cV})$ given with respect to the preferred frame $S_V$ (Fig.1), so that the momentum is non-null ($P_0$) and the energy is equivalent to the rest energy $m_0c^2$ in SR with $p=0$ for $v=0$. However, in SSR, $E_0$ is associated with the speed $v_0$ (reference frame $S_0$) with respect to the ultra-referential $S_V$, as there is no rest in the space-time of SSR, where the references frames $S$, $S^{\prime}$ and specially $S_0$ for $v=v_0(=\sqrt{cV})$ are shown in Fig.1. 

The SSR-metric is a deformed Minkowski metric with the presence of the multiplicative factor $\Theta=\Theta(v)=1/(1-V^2/v^2)$\cite{Rodrigo}, which plays the role of a conformal factor as already shown in a previous paper\cite{Rodrigo}, thus leading to a Conformal Special Relativity represented by SSR due to the presence of the invariant minimum speed $V$, as follows: 

\begin{equation}
d\mathcal S^{2}=\frac{1}{\left(1-V^2/v^2\right)}[c^2(dt)^2-(dx)^2-(dy)^2-(dz)^2],
\end{equation}
or simply $d\mathcal S^{2}=\Theta\eta_{\mu\nu}dx^{\mu}dx^{\nu}$, where $\Theta=1/(1-V^2/v^2)$ and $\eta_{\mu\nu}$ is the Minkowski metric. 

By dividing Eq.(1) by Eq.(2), we obtain the following velocity transformation in SSR, namely: 

\begin{equation}
v_{rel}=v_{S'/S}=\frac{v^{\prime}-v+V}{1-\frac{v^{\prime}v}{c^2}+\frac{v^{\prime}V}{c^2}}=\frac{v^{\prime}-v(1-\alpha)}{1-\frac{v^{\prime}v(1-\alpha)}{c^2}}, 
\end{equation}
where $\alpha=V/v$.

We have considered $v_{rel}=v_{relative}=v_{S'/S}\equiv dx^{\prime}/dt^{\prime}$ and $v^{\prime}\equiv dX/dt$ when dividing Eq.(1) by Eq.(2). 

We should stress that $v^{\prime}=v_{S'}\equiv dX/dt$ is the motion of the referential $S^{\prime}$ (Fig.1) with respect to the background reference frame $S_V$ connected to the unattainable minimum speed $V$, i.e., we can write the notation $v_{S'}=v_{S'/S_V}$ for representing the absolute motion of $S^{\prime}$, which is observer-independent, as $S_V$ is absolute for being the preferred reference frame.  

The speed $v$ shown in Fig.1 represents the motion of the referential $S$ with respect to the background reference frame $S_V$, i.e., we can write the notation $v=v_{S}=v_{S/S_V}$ for representing the absolute motion of $S$ (Fig.1), which is also observer-independent.

The speed $v_{rel}$ is the relative speed between the absolute speeds $v_{S'}$ and $v_{S}$, both of them given in relation to the background framework $S_V$, i.e., we have $v_{rel}=v_{S'/S}$ (Fig.1). So we can rewrite the speed transformation in 
Eq.(4) by using the notations with the presence of the background frame $S_V$, namely: 

\begin{equation}
v_{rel}=v_{S'/S}=\frac{v_{S'/S_V}-v_{S/S_V}+V}{1-\frac{(v_{S'/S_V})(v_{S/S_V})}{c^2}+\frac{(v_{S'/S_V})V}{c^2}}, 
\end{equation}
where $v=v_{S}=v_{S/S_V}$ (speed $v$ of the reference frame $S$ in relation to $S_V$) and $v'=v_{S'}=v_{S'/S_V}$ (speed $v^{\prime}$ of the reference frame $S^{\prime}$ in relation to $S_V$).  

Fig.1 also shows the reference frame $S_0$, whose speed $v_0(=\sqrt{cV})$ is also given in relation to $S_V$. 

As $v_0$ is the intermediary speed, such that $V<<v_0<<c$ with $\Psi(v_0)=\Psi(\sqrt{cV})=1$, all the speeds $v$ not so far from $v_0$, where $E\approx E_0=m_0c^2$ represent the Newtonian approximation within the scenario of SSR, as we get $\Psi(V<<v<<c)\approx 1$. 

If $V\rightarrow 0$, Eq.(5) would recover the Lorentz velocity transformation, where both speeds $v_{S'}$ and $v_S$ would be given simply in relation to a certain Galilean frame at rest in lab, such that the background frame $S_V$ would vanish and thus $v_0$ would be also zero, i.e., the reference frame $S_0$ would become simply a certain Galilean reference frame at rest in lab. 

From the transformation in Eq.(5), let us just consider the important cases, where we must consider $v_{S'}\geq v_{S}$ (Fig.1), namely: 

 {\bf a)} If $v_{S'}=c$ (photon) and $v_S\leq c$, this implies in $v_{rel}=c$. Such result just verifies the invariance of $c$.

 {\bf b)} If $v_{S'}>v_S(=V)$, this implies in $v_{rel}=``v_{S'}-V"=v_{S'}$. For example, if $v_{S'}=2V$ and $v_S=V$, this leads to $v_{rel}=``2V-V"=2V$, which means that $V$ ($S_V$) really has no influence on the speeds of any particles. Thus, $V$ works as if it were an ``absolute zero of motion'', being invariant and having the same value at all directions of space $3D$ of the isotropic background field associated with $S_V$. 

 {\bf c)} If $v_{S'}=v_S$, this implies in $v_{rel}=v_{S'/S}=``v_S-v_S"=``v-v"$($\neq 0)=\frac{V}{1-\frac{v_S^2}{c^2}(1-\frac{V}{v_S})}=
 \frac{V}{1-\frac{v^2}{c^2}(1-\frac{V}{v})}$. 

From the case ({\bf c}), let us consider two specific cases, as follows:

-$c_1$) Assuming that $v_S=V$, this implies in $v_{rel}=``V-V"=V$ as verified before. Indeed $V$ is an invariant minimum speed. 

-$c_2$) If $v_S=c$ (photon), this implies in $v_{rel}=c$, where we have the interval $V\leq v_{rel}\leq c$ given for the interval $V\leq v_S\leq c$. However, it must be stressed that there is no ordinary massive particle exactly at the ultra-referential $S_V$ with $v=v_S=V$. So, this is just a hypothetical condition to verify the consistency of the transformation in Eq.(5) with respect to the invariance of the minimum speed $V$, as already verified in the specific case $c_1$. 

This last case ({\bf c}) shows that it is impossible to find the rest for the particle on its own reference frame $S$, where $v_{rel}(v_S)$ ($\equiv\Delta v(v_S)$) is a function that increases with the increasing of $v=v_S$ of the referential $S$ (Fig.1). However, if we make $V\rightarrow 0$, so we would have $v_{rel}\equiv\Delta v=0$ and thus it would be possible to find rest for $S$, which would recover the inertial reference frames of SR.

The inverse transformations of space-time ($x^{\prime}\rightarrow X$) and ($t^{\prime}\rightarrow t$) in SSR for the special case $(1+1)D$ and also the general case $(3+1)D$ have already been explored in details in a previous paper\cite{N2016}. Thus, from such transformations above, we can obtain the following inverse transformation of velocity, namely: 

\begin{equation}
v_{rel}=v_{S'/S}=\frac{v_{S'/S_V}+v_{S/S_V}-V}{1+\frac{(v_{S'/S_V})(v_{S/S_V})}{c^2}-\frac{(v_{S'/S_V})V}{c^2}}. 
\end{equation}

The velocity transformation given by Eq.(6) leads to the following important cases:  

 {\bf a)} If $v^{\prime}=v_{S'}=v_S=v=V$, this implies in $``V+V"=V$. Once again we verify that the minimum speed $V$ is in fact invariant. 
 
 {\bf b)} If $v^{\prime}=v_{S'}=c$ (photon) and $v_S\leq c$, this leads to $v_{rel}=v_{S'/S}=c$. This just confirms that $c$ is invariant. 
 
 {\bf c)} If $v^{\prime}=v_{S'}>V$ and by considering $v_S=V$, this leads to $v_{rel}=v_{S'/S}=v_{S'}$. 
 
 From the case ({\bf c}), let us investigate the following specific cases, namely: 
 
 -$c_{1}$) If $v^{\prime}=v_{S'}=2V$ and by assuming that $v_S=V$, we would obtain $v_{rel}=v_{S'/S}=``2V+V"=2V$. 
 
 -$c_{2}$) If $v^{\prime}=v_{S'}=v_S=v$, this implies in $v_{rel}=v_{S'/S}=``v_S+v_S"=``v+v"=\frac{2v_S-V}{1+\frac{v_S^2}{c^2}(1-\frac{V}{v_S})}=\frac{2v-V}{1+\frac{v^2}{c^2}(1-\frac{V}{v})}$.
 
 In the Newtonian regime ($V<<v<<c$) for $c_2$ above, we recover the classical transformation, i.e., $v_{rel}=``v+v"=2v$. 
 
 In the relativistic regime ($v\rightarrow c$), we recover the Lorentz transformation of velocity given for this specific case $c_2$ ($v^{\prime}=v_{S'}=v_S=v$), i.e., we find 
 $v_{rel}=v_{S'/S}=``v_S+v_S"=``v+v"=2v_S/(1+v_S^2/c^2)=2v/(1+v^2/c^2)$.
 
 \section{\label{sec:level1} Equivalence of the SSR-metric with a dS-metric}

\begin{figure}
\begin{center}
\includegraphics[scale=0.27]{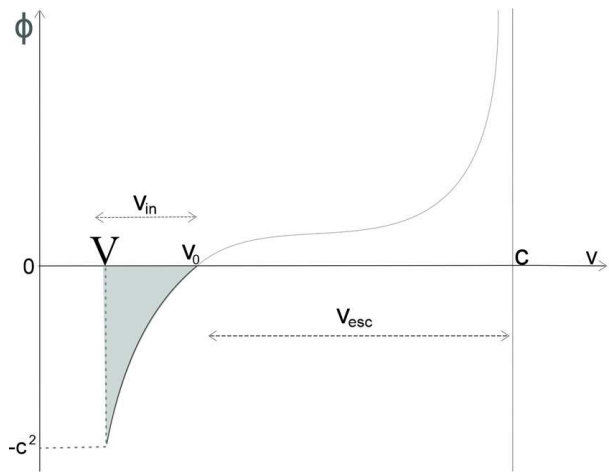}
\end{center}
\caption{This figure shows the scalar potential $\phi(v)=c^{2}\left(\sqrt{\frac{1-\frac{V^2}{v^2}}{1-\frac{v^2}{c^2}}}-1\right)$ given in function of speed [Eq.(8)]. It shows two phases, namely gravity (right side)/anti-gravity (left site), where the barrier at the right side represents the relativistic limit, i.e., speed of light $c$ with $\phi\rightarrow\infty$, and on the other hand, the barrier at the left side is the anti-gravitational limit only described by SSR with an invariant minimum speed $V$ associated with the potential $\phi_q(V)=\phi(V)=-c^2$. The intermediary region is the Newtonian regime ($V<<v<<c$), where occurs a phase transition of gravity/anti-gravity for $v=v_0=\sqrt{cV}$.}
\end{figure}
 
As the universal minimum speed related to the ultra-referential $S_V$ shoud be associated with the cosmological constant, let us show that the metric of SSR [Eq.(3)] is equivalent to a dS-metric, where there emerges a conformal factor depending on $\Lambda$\cite{Rodrigo}. To do that, let us use a model by considering a spherical universe with Hubble radius $R_H$ filled by a vacuum energy density $\rho_{\Lambda}$. 
 
According to this model, on the surface of the sphere by representing the frontier of the observable universe, all the objects like galaxies, etc experience an anti-gravity given by the accelerated expansion of the universe. This anti-gravitational effect is due to the whole vacuum energy (a dark mass) inside such a Hubble sphere. Thus, we think that each galaxy works like a proof body that interacts with this sphere having a dark mass $M_{\Lambda}(=M)$. Such interaction can be thought of as being the simple case of interaction between two bodies. In view of this, let us show that there is an anti-gravitational interaction between the ordinary proof mass $m_0$ (on the surface of the dark sphere) and the own dark sphere with a dark mass $M$. 

So, in order to investigate such anti-gravitational interaction between the proof mass $m_0$ and the dark mass $M$ of the Hubble sphere with radius $R_H$ in this model, let us first remind the model of a proof particle with mass $m_0$ that escapes from a gravitational potential $\phi$ on the surface of a certain sphere of matter with mass $M_{matter}$, namely $E=m_0c^2(1-v^2/c^2)^{-1/2}\equiv m_0c^2(1+\phi/c^2)$, where $E$ is the escape relativistic energy of the proof particle with mass
$m_0$ and $\phi=GM_{matter}/R$, $R$ being the radius of the sphere of matter. In this classical case, the interval of escape velocity ($0\leq v<c$) is associated with the interval of potential ($0\leq\phi<\infty$), where we define $\phi>0$ to be the well-known classical (attractive) gravitational potential. 

We should notice that the Lorentz symmetry violation in SSR is due to the presence of the ultra-referential $S_V$ (Fig.1) connected to the vacuum  energy that fills the dark sphere. Such energy has origin from a non-classical aspect of gravity that leads to a repulsive gravitational potential defined as being negative for representing anti-gravity, i.e., 
$\phi=\phi_q<0$ (Fig.2). 

In this model of spherical universe based on SSR-theory, we write the deformed relativistic energy of such a proof particle ($m_0$), as follows: 

\begin{equation}
E=m_0c^2\left(1+\frac{\phi}{c^2}\right)=
m_0c^2\left(\frac{{1-\frac{V^2}{v^2}}}{{1-\frac{v^2}{c^2}}}\right)^{\frac{1}{2}},
\end{equation}
from where we get
\begin{equation}
\phi=\phi(v)=\left[\left(\frac{{1-\frac{V^2}{v^2}}}{{1-\frac{v^2}{c^2}}}\right)^{\frac{1}{2}}-1\right]c^2. 
\end{equation} 

Here, we should realize that Eq.(8) reveals two situations, namely: 

i) the well-known Lorentz sector 
($\phi=(\gamma-1)c^2$) represents the gravity sector, since the sphere $M$ is composed by attractive (ordinary) matter. In this case, the speed $v$ is simply the escape velocity ($v_{esc}$), which is directed away from the sphere.

ii) the anti-gravity sector 
($\phi=\phi_q=(\theta-1)c^2$),
where $\theta=(1-V^2/v^2)^{1/2}$ is governed by a dark sphere with mass $M$. Here, the speed $v$ is the input speed ($v_{in}$) or the velocity of a proof particle that escapes from anti-gravity, i.e., $v(=v_{in})$ is directed into the sphere, since anti-gravity pushes the particle away. 

As SSR forbids rest of a particle according to Eq.(8), we must be careful to notice that $v$ cannot be zero even in the absence of potential $\phi$ ($\phi=0$), i.e, we find $v=v_0=\sqrt{cV}$, so that $\phi(v_0)=0$ 
in Eq.(8) (Fig.2). 

Due to the absence of gravitational potential 
($\phi=0$) at the point $v=v_0(\neq 0)$, this is the only velocity that means both of the escape and input velocities of a particle. Therefore, $v_0$ is a zero-point transition between gravity and anti-gravity, which highlights the quantum nature of the space-time in SSR, thus leading to the uncertainty principle as shown in a previous paper\cite{uncertainty}. 

In short, from Eq.(8) and Fig.(2) we can see two regimes of gravitational potential, i.e., the classical (matter) and quantum (vacuum) regimes, namely: 

\begin{equation}
\phi=\phi(v)=\left\{
\begin{array}{ll}
\phi_{q}:&\mbox{$-c^2<\phi\leq 0$ for $V< v\leq v_0$}.\\\\
\phi_{m}:&\mbox{$0\leq\phi<\infty$ for $v_0\leq v<c$}, 
\end{array}
\right.
\end{equation}
so that the speed $v_0$ represents the point of transition ($\phi=0$) between gravity 
($\phi_{matter}=\phi_{m}=\phi>0$ for $v>v_0$) and anti-gravity when the vacuum governs 
($\phi_{quantum}=\phi_q=\phi<0$ for $v<v_0$). 

We must stress that $v_0$ is given with respect to the preferred reference frame $S_V$ (Fig.1). Therefore, it is an observer-independent velocity as well as any velocity $v$, which is given with respect to $S_V$, since $S_V$ is related to the unattainable minimum speed $V$, and thus there is no observer at the ultra-referential $S_V$.   

We realize that the most repulsive potential is 
$\phi=-c^2$, which is associated with the fundamental vacuum energy of the ultra-referential $S_V$ by imposing $v=v_{in}=V$ in Eq.(8), i.e., $\phi(V)=-c^2$ (Fig.2). 

So, by taking into account this model of a spherical universe with a Hubble radius $R_H(=R_u)$ and a vacuum energy density $\rho_{\Lambda}$, we obtain the total vacuum energy inside the sphere, i.e., $E_{dark}=\rho V_u=Mc^2$, where  $V_u$ is the spherical volume of the universe
and $M$ is the total dark mass associated with the vacuum energy inside the sphere. 

As the vacuum energy density $\rho_{\Lambda}$ is very low and the big sphere with Hubble radius 
$R_H(=R_u)$ presents a dark mass $M$, but having a very low dark mass density, then the Newtonian gravitational potential is a very good approximation that represents this toy model for the universe. So, in view of this, we get the following repulsive gravitational potential $\phi(=\phi_q<0)$ on the surface of such Hubble sphere (universe), namely: 

\begin{equation}
\phi=\phi_q=-\frac{GM}{R_u}=
-\frac{4\pi G\rho R_u^2}{3c^2}=-\frac{G\rho V_u}{R_uc^2},
\end{equation}
where $M=\rho V_u/c^2$, $\rho=\rho_{\Lambda}$ is the vacuum energy density and $V_u(=4\pi R_u^3/3)$ is the Hubble volume.

We already know that $\rho=\Lambda c^2/8\pi G$. So, by substituting this relationship ($\rho$) in Eq.(10), we obtain the repulsive (quantum) potential, as follows: 

\begin{equation}
\phi=-\frac{\Lambda R_u^2}{6},
\end{equation}
where $R_u=R_H=cT_H(\sim 10^{26}$m) is the Hubble radius. $T_H\cong 13.7$ Gyear is the age of the universe (Hubble time). 

As the whole universe (a big sphere) is governed by the vacuum energy (a dark mass $M$), the speed $v$ in Eq.(8) is understood as the input speed $v_{in}$ in order to overcome the cosmological anti-gravity. Thus, the factor
$(1-V^2/v^2)^{1/2}$ [Eq.(8)] prevails for determining the potential $\phi$. In view of this, we will neglect the Lorentz factor $\gamma$ (attractive sector) in Eq.(8), and so we should consider only the repulsive sector ($V<v\leq v_0$) for obtaining the non-classical background potential $\phi(=\phi_q)$.

Furthermore, after neglecting $\gamma=
(1-v^2/c^2)^{-1/2}$ in Eq.(8), we will just compare its anti-gravity sector with Eq.(15) given for a certain radius $r(=ct)$, i.e., $\phi(=-\Lambda r^2/6)$, so that we find 
$\phi/c^2$, namely: 

\begin{equation}
\frac{\phi}{c^2}=-\frac{\Lambda r^2}{6c^2}=\left(1-\frac{V^2}{v^2}\right)^{\frac{1}{2}}-1, 
\end{equation}
where $\phi=\phi_q$, which represents the potentials of anti-gravity (Fig.2), being $0\leq\phi\leq-c^2$. 

By performing the calculations in Eq.(12), we rewrite the scale factor $\Theta(v)$ of the SSR-metric [Eq.(3)] in its equivalent forms, as follows:

\begin{equation}
\Theta(v)=\frac{1}{\left(1-\frac{V^2}{v^2}\right)}\equiv\frac{1}{\left(1+\frac{\phi_q}{c^2}\right)^2}\equiv\frac{1}{\left(1-\frac{\Lambda r^2}{6c^2}\right)^2}, 
\end{equation}
where we realize that there are three equivalent forms for representing $\Theta(v)\equiv\Theta(\phi_q)\equiv\Theta(\Lambda)$ as shown in Eq.(13). 

By replacing the factor $\Theta(v)$ of Eq.(3) (SSR-metric) by its equivalent form with dependence of $\Lambda$ shown in Eq.(13), we rewrite the background metric (SSR-metric) in its equivalent form within the dS-scenario, namely: 

\begin{equation}
d\mathcal S^{2}=\frac{1}{\left(1-\frac{{\Lambda}r^2}{6c^2}\right)^2}[c^2(dt)^2-(dx)^2-(dy)^2-(dz)^2], 
\end{equation}

or simply 

\begin{equation}
d\mathcal S^{2}=\Theta(\Lambda)\eta_{\mu\nu}dx^{\mu}dx^{\nu}, 
\end{equation}
where $\eta_{\mu\nu}$ is the Minkowski metric and
$\mathcal G_{\mu\nu}=\Theta(\Lambda)
\eta_{\mu\nu}$ is the SSR-metric with dependence of $\Lambda$. 

Of course if we make $\Lambda=0$ in Eq.(14), we get $\Theta=1$ and so we recover the Minkowski metric $\eta_{\mu\nu}$, where there is no cosmological constant and no anti-gravitational effect. In other words, as 
$\Lambda=-6\phi/r^2$ [Eq.(11)], 
for $r\rightarrow\infty$ 
($\Lambda\rightarrow 0$), the interval 
$d\mathcal S^{2}$ reduces to the 
Lorentz-invariant Minkowski interval $ds^2$, i.e., $d\mathcal S^{2}\rightarrow
ds^2=\eta_{\mu\nu}dx^{\mu}dx^{\nu}$
\cite{jgpereira}. 

We should realize that Eq.(14) represents a 
dS-metric which presents $\Lambda>0$, as we
must have $\phi<0$ (anti-gravity sector) according to Eq.(11).   

In view of Eq.(11) and Eq.(14), we first conclude that a cosmological constant $\Lambda$ emerges from SSR, i.e., $\Lambda=-6\phi/r^2$ [Eq.(11)]. We also conclude that there is a correspondance of SSR with the de-Sitter (dS) relativity\cite{jgpereira} shown by Eq.(14) that is a dS-metric with a conformal factor\cite{jgpereira} given by $\Theta(\Lambda)$. 

We finally conclude that a small positive value of $\Lambda$ is plausible in the context of Eq.(11) and Eq.(14). So, in order to estimate the small order of magnitude of $\Lambda$, we first consider $\Lambda$ [Eq.(11)] given for the Hubble radius $R_H(\sim 10^{26})m$, so that we obtain 

\begin{equation}
\Lambda=\Lambda(R_H,\phi)=-\frac{6\phi}{R_H^2},
\end{equation}
where $r=R_H$ and $-c^2\leq\phi\leq 0$ (the shaded area in Fig.2). 

Finally, if we admit that the accelerated expansion of the universe is governed by the lowest potential $\phi=\phi(V)=-c^2$ associated with the fundamental vacuum energy at the ultra-referential $S_V$, we find 

\begin{equation}
\Lambda=\frac{6c^2}{R_H^2}\sim 10^{-35}s^{-2}. 
\end{equation} 

\subsection{\label{sec:level1} The cosmological constant in the zero-gravity limit ($z=1$) according to the Boomerang experiment} 

The very small $\Lambda(\sim 10^{-35}s^{-2})$ may have implication in a realistic cosmological scenario of a flat space-time governed by a dark energy ($\Omega_{\Lambda}\approx 0.7$) according to the Boomerang experiment\cite{boomerang}\cite{boomerang1}, which 
is consistent with the $\Lambda CDM$ scenario. In order to realize such an implication, we first need to approximate the metric given in Eq.(14) to a quasi-flat metric representing a universe with a slightly accelerated expansion, so that we should make $\Lambda\approx 0$ or even $\Lambda<<1$ in Eq.(14). In doing that, we get the approximation for the case of a very weak anti-gravity, i.e., we make the approximation $\Lambda r^2/6<<c^2$ in Eq.(14). Thus, we write $\Theta=(1-\Lambda r^2/6c^2)^{-2}\approx(1-\Lambda r^2/3c^2)^{-1}$, so that we obtain the following metric: 

\begin{equation}
d\mathcal S^{2}=\frac{1}{\left(1-\frac{{\Lambda}r^2}{3c^2}\right)}[c^2(dt)^2-(dx)^2-(dy)^2-(dz)^2], 
\end{equation}
from where we can get the cosmological constant within the scenario 
of a very weak anti-gravity. In order to do that, we consider the most fundamental vacuum at $S_V$ (Fig.1), so that we make $\phi=-c^2$ in Eq.(16). Thus we obtain $\Lambda$ in function of the Hubble time, namely: 

\begin{equation}
\Lambda=3c^2/r^2=3/\tau^2, 
\end{equation}
with $r=c\tau$, $\tau$ being a certain Hubble time. 

We know that $\Lambda=8\pi G\rho/c^2=k\rho$, where $8\pi G/c^2$ is 
the well-known constant $k$ in the Einstein equation. So we can write
$k\rho=3/\tau^2$, where we find $\rho=3/k\tau^2=\Lambda/k$.

Now, we stress that there must be a critical density $\rho_c=3/k\tau_0^2$ given exactly in the zero-gravity limit associated with the Hubble time 
$\tau_0$ at which the universe goes over from a decelerating to an accelerating expansion. Thus the obtaining of the numerical value $\tau_0$ will allow us to get the numerical value of $\Lambda_0(=k\rho_c=3/\tau_0^2)$ in agreement with measurements\cite{boomerang}\cite{boomerang1}. 

Here we must quote a previous paper entitled ``Fundamental Approach to the
Cosmological Constant Issue''\cite{Lambda}, where one confirms that the
universe now is definitely in a stage of accelerating expansion. Although this theory\cite{Lambda} has no cosmological constant, it predicts that the universe accelerates and hence it has the equivalence of a positive cosmological constant in Einstein's general relativity. In the framework of this theory\cite{Lambda} the zero-zero component of the field equations ($R_{\mu\nu}-(R/2)g_{\mu\nu}=kT_{\mu\nu}$) is written as 

\begin{equation}
R^0_0-\frac{1}{2}\delta^0_0 R=k\rho_{eff}=k(\rho-\rho_c),
\end{equation}
where $\rho_c=3/k\tau_0^2=\Lambda_0/k$\cite{Lambda} is exactly the critical density and $\tau_0$ is Hubble's time in the zero-gravity limit.

In Eq.(20), it is important to note that the effective density is null ($\rho_{eff}=0$) only if $\rho=\rho_c$, which corresponds exactly to the zero-gravity limit. 

Here we must stress that the framework of this theory\cite{Lambda} given 
by Eq.(20) is consistent with Eq.(18), since both equations provide information about the existence of a critical density $\rho_c$ and a Hubble time $\tau_0$ at the zero-gravity limit. Such consistency is not surprising, since in a previous paper\cite{Rodrigo} it has already been proven that the conformal metric of SSR is a solution of the Einstein equation in dS-scenario ($\Lambda>0$).

The framework that leads to Eq.(20)\cite{Lambda} uses a Riemannian 
four-dimensional presentation of gravitation in which the coordinates are those of Hubble, i.e., distances and velocity rather than the traditional space and time. When solving the field equations, there are three possibilities for the universe to expand, namely a decelerating
expansion ($\Omega_m>1$), followed by a constant expansion 
($\Omega_m=1$) and after an accelerating expansion ($\Omega_m<1$), 
and it predicts that the universe is now in the latter phase.

As we are interested only in the latter phase of acceleration described by the background dS-metric in Eq.(18), then for the accelerating phase given by the theory\cite{Lambda}, we find  

\begin{equation}
 H_0=h[1-(1-\Omega_m)z^2/6], 
\end{equation}
where $\Omega_m=\rho_m/\rho_c(<1)$ and $h=\tau^{-1}$.  

The redshift parameter $z$ determines the distance at which 
$H_0$ is measured. So, at the zero-gravity limit, by choosing $z=1$ and take for $\Omega_m= 0.245$, its value at the present time (see the 
Table 1 in ref.\cite{Lambda}), Eq.(21) then gives 
$H_0=0.874h$. 

For $z=1$, the corresponding Hubble parameter $H_0$ can be taken as 
$H_0=70km/s-Mpc$, and thus $h=h_0=(70/0.874)km/s-Mpc$, or 
$h_0=80.092km/s-Mpc$, and so we get $\tau=\tau_0=12.486 Gyr=3.938\times 10^{17}s$, which is exactly the numerical value of the Hubble time 
$\tau_0$ at which the universe goes over from a decelerating to an accelerating expansion. 

 By substituting the numerical value of $\tau_0$ above in 
 Eq.(19), we obtain 
 
 \begin{equation}
 \Lambda_0=\frac{3}{\tau_0^2}=1.934\times 10^{-35}s^{-2}.
 \end{equation}
 
 In sum we realize that the SSR-theory has proposed a new structure of space-time with an invariant minimum speed $V$ associated with the preferred reference frame $S_V$ for representing the vacuum energy. Therefore SSR amplifies the framework of special relativity (SR), which leads to provide the first principles for understanding the tiny positive value of the cosmological constant in excellent agreement with the observational data\cite{boomerang}\cite{boomerang1}
 \cite{15,16,17,18,19,20,21} by avoiding the renormalization procedures of the quantum vacuum in the Quantum Field Theories (QFT), which lead to the well-known {\it Cosmological Constant Problem}\cite{Lambda1}\cite{Lambda2}\cite{Lambda3}. 
 
\section{The Weyl Geometrical Structure of SSR}
 
An important aspect of Weyl geometric structure is that when one goes from one frame $(M, g, \sigma)$ to another frame $(M, \bar{g}, \bar{\sigma})$ by using the gauge transformations \cite{Rosen}\cite{Romero}, we have
 
\begin{equation} \label{Weyl}
\begin{aligned}
\bar{g} &= e^{f}g \\
\bar{\sigma} &= \sigma + df, 
\end{aligned}
\end{equation}
the affine geodesic curves, where $f$ is a scalar function defined in the differentiable manifold $M$, with a metric $g$ and the Weyl field $\sigma$ one keeps unaltered. In a certain sense we can characterize the Weyl geometry as an extension of Riemannian geometry where the covariant derivative of the Metric Tensor $g$ is given by 

\begin{equation}\label{affine}
\nabla_\alpha g_{\beta \lambda} = \sigma_{\alpha}g_{\beta \lambda},
\end{equation}
with $\sigma_\alpha$ being the components of a one-form field $\sigma$ in a local coordinate basis. In this case we can write the affine connection as\cite{Romerob}

\begin{equation}
\Gamma^{\alpha}_{\beta \lambda} = \left\{ \begin{array}{c} k \\ ij \end{array} \right\} -\frac{1}{2}g^{\alpha \mu}[g_{\mu\beta} \sigma_{\lambda} + g_{\mu\lambda} \sigma_{\beta}-g_{\beta \lambda} \sigma_\mu],
\end{equation}
where $\left\{ \begin{array}{c} k \\ ij \end{array} \right\}$ is the usual Christoffel symbols.\\

Thus the equations of motion, which are the field or geodesic equations of this geometric structure

\begin{equation}\label{geo}
\frac{d^2 x^{\alpha}}{d\tau^2} + \Gamma^{\alpha}_{\beta \lambda}\frac{dx^{\beta}}{d\tau}\frac{dx^{\lambda}}{d\tau} = 0
\end{equation}
are invariant, with $\tau$ being the usual proper time of the field equations of general relativity.

The beauty of Weyl's treatment is that under the condition (\ref{affine}) the connection $\nabla_{\alpha}$ and therefore the geodesic equations are invariant under transformations (\ref{Weyl}). On the other hand consider two vector fields $V$ and $U$ which can be parallel-transported along a curve $\alpha = \alpha(\lambda)$. So from (\ref{affine}) one can written

\begin{equation}\label{aaffine}
\frac{d}{d\lambda}g(V, U)=\sigma(\frac{d}{d\lambda})g(V,U), 
\end{equation}
with $\frac{d}{d\lambda}$ being the vector tangent to $\alpha$. Integrating (\ref{aaffine}) along the curve $\alpha$ from a point $P_0 = \alpha(\lambda_0)$ one obtains

\begin{equation}\label{iaaffine}
g(V(\lambda), U(\lambda)) = g(V(\lambda_0), U(\lambda_0))e^{\int^{\lambda}_{\lambda_0}\sigma(\frac{d}{d\rho})\rho}. 
\end{equation}

For a specific case where $U = V$ and $L(\lambda)$ is the length of the vector $V(\lambda)$ at a point $P = \alpha(\lambda)$ of the curve, so in the coordinate system ${x^{\alpha}}$, the relation (\ref{aaffine}) becomes 

\begin{equation}\label{length}
\frac{dL}{d\lambda} = \frac{\sigma_\alpha}{2}\frac{dx^{\alpha}}{d\lambda}L.
\end{equation}

Considering a closed curve with the conditions $\alpha \ : [a,b] \ \in \ R \rightarrow M  $ or in a concise way $\alpha(a) = \alpha(b)$ the both concepts (\ref{iaaffine}) and (\ref{length}) conduct to

\begin{equation}\label{flength}
L = L_0e^{\frac{1}{2}\oint \sigma_\alpha dx^{\alpha}}, 
\end{equation}
with $L_0$ and $L$ being respectively the values of $L(\lambda)$ at $a$ and $b$.

The theoretical consequence of (\ref{flength}) is that applying the Stokes theorem we obtain the expression 

\begin{equation}
L = L_0e^{-\frac{1}{4}}\int \int F_{\mu \nu}dx^{\mu} \wedge dx^{\nu},
\end{equation}
so that now we can define $F_{\mu\nu} = \partial_{\nu}\sigma_{\mu}-\partial_{\mu}\sigma_{\nu}$. The result that one obtains is that a vector have its original length preserved if the 2-form $F = d\sigma = \frac{1}{2} F_{\mu\nu} dx^{\nu} \wedge dx^{\mu}$ vanishes.
\
\subsection{\label{sec:level1} Weyl Conformally Flat Spacetimes and Gravity}

We wish to show in a simple and direct way that the factor $\Theta(v)$ in Eq.(3) is a conformal Weyl factor and that the SSR-theory has a Weyl conformal geometry by verifying its Newtonian weak-field limit in the scenario of SSR, i.e., $V<<v<<c$, namely we have the approximation

\begin{equation}
g_{\mu\nu}\approx\eta_{\mu\nu} + \epsilon h_{\mu\nu}
\end{equation},
where $\eta_{\mu\nu}$ is the Minkowski tensor, $\epsilon$ is a small parameter and $\epsilon h_{\mu\nu}$ is a small time-independent perturbation on the metric tensor 
$g$. 

Now by considering a conformally flat spacetime we can write

\begin{equation}
g_{\mu\nu} = e^{\phi}\eta_{\mu\nu}\backsimeq (1 + \phi)\eta_{\mu\nu},
\end{equation}
where it is interesting to observe that for $\Lambda<<1$, which means $\phi<<1$ ($c=1$) or $V<<v<<c$ (Newtonian limit in SSR) in Eq.(3), this leads to the aproximation $g_{\mu\nu}\approx (1+\phi)\eta_{\mu\nu}\approx{(1+\Lambda r^2/3)}\eta_{\mu\nu}$ ($\Lambda<<1$ ; $c=1$), which is valid for redshift $z\cong 1$ with a very weak antigravity (almost flat space-time) with a slight cosmic acceleration, from where we have obtained the very small value of $\Lambda=\Lambda_0=1.934\times 10^{-35}s^{-2}$ in the approximation of Euclidian space or zero-gravity (quasi-flat space). 

Therefore, in this regime of transition from gravity (zero-gravity) to anti-gravity, the Minkowski metric becomes slightly deformed, namely:

\begin{equation}\label{Min}
ds^2\approx (1 + \phi)[(dx^0)^2-(dx^1)^2]-(dx^2)^2-(dx^3)^2],
\end{equation}
with $x^0 = ct$ and $\phi<<1$ (almost flat space-time). 

If we consider the motion of a test particle in the spacetime (\ref{Min}), as 
$\Gamma^{\mu}_{\alpha \beta}$ is invariant under (\ref{Weyl}), the following approximation can be realized

\begin{equation}\label{approximation}
\left(\frac{ds}{dt}\right)^2 \backsimeq c^2(1 + \epsilon h_{00}) = c^2(1 + \phi).
\end{equation}

So the geodesic equation is

\begin{equation}
\frac{d^2x^{\mu}}{ds^2} + \Gamma^{\mu}_{\alpha\beta}\frac{dx^{\alpha}}{ds}\frac{dx^{\beta}}{ds}, 
\end{equation}
which reduces itself to 

\begin{equation}
\frac{d^2x^{\mu}}{dt^2} + \Gamma^{\mu}_{00}\left(\frac{dx^0}{ds}\right)^2, 
\end{equation}
which in the approximation (\ref{approximation}) may be written as 

\begin{equation}
\frac{d^2x^{\mu}}{dt^2} + c^2\Gamma^{\mu}_{00}
\end{equation}

Thus we can observe that the conformal factor in Eq.(3) includes the same extended geometrical structure and the same weak-field Newtonian limit, i.e., $V<<v<<c$ or 
$\phi<<1$ for $c=1$. 

It is worth mentioning that the Weyl structure is overlying to Riemannian geometry and the fact that the field equations are invariant under the frame transformations occasioned by the transformations (\ref{Weyl}), in a certain proportion show that the geometric structure of the SSR can address fundamental aspects of nature such as the nature of the cosmological constant and dark energy.
 
\section{\label{sec:level1} Conclusions and prospects} 

First of all, based on the quantum principle of zero-point energy that originates from the uncertainty principle, which is not consistent with the classical space-time of Special Relativity (SR), we have searched for a quantum space-time structure being consistent with the idea of the absence of rest in the quantum world by precisely postulating an invariant minimum speed $V$, which allowed us to include the concept of vacuum associated with a preferred reference frame $S_V$ (Fig.1). 

The minimum speed is a new kinematic invariant given for lower energies, which led to a new Deformed Special Relativity (DSR) so-called Symmetrical Special Relativity (SSR), from where there emerged the cosmological constant, which allowed us to show the equivalence of SSR-metric with a dS-metric. The small order of magnitude of the cosmological constant ($\Lambda\sim 10^{-35}s^{-2}$) was estimated. 

Finally, we were able to obtain the tiny numerical value of the cosmological constant $\Lambda_0=1.934\times 10^{-35}s^{-2}$ given in the zero-gravity limit (flat universe) when considering the redshift $z=1$ and the Hubble time $\tau_0$ at which the universe goes over from a decelerating to an accelerating expansion. 

Section 5 was dedicated to the Weyl geometrical structure of SSR. We have shown that the factor $\Theta(v)$ in Eq.(3) behaves like a conformal Weyl factor so that SSR includes a Weyl conformal geometry in the regime of Newtonian weak-field given in the SSR scenario 
($V<<v<<c$). Such regime corresponds to a slight acceleration of the universe given for redshift $z=1$, where we have obtained the tiny value of the cosmological constant according to the experiments. So we have concluded that the current expanding universe is governed by a Weyl conformal geometry for weak-field ($\phi<<1$) by representing an almost flat space-time as a special case of Eq.(3) of SSR. 

Here it is important to call attention to a chapter of a book that 
was the first publication that considered the connection between the minimum speed and the Weyl geometry\cite{Weyl1}. 

Another important work\cite{acoustic} has considered the relation of the minimum speed with superfluids and acoustic geometries and 
the latter is a review published in AIP Advanced, which makes a review about Relativistic Hydrodynamics\cite{Hydrodynamics}. 

The investigation of symmetries of the SSR-theory should be done by means of their association with a new kind of electromagnetism when we are in the limit $v\rightarrow V$, which could explain the problem of high magnetic fields in magnetars\cite{gravastar}\cite{gravastar1}\cite{Greiner}, super-fluids in the interior of gravastars\cite{gravastar1} and other types of black hole mimickers\cite{haw}\cite{Tds,Tds1,haw2}.\\\\

{\bf Data Availability}

Data sharing not applicable to this article as no datasets were generated or analysed during the current study. \\

\end{document}